\documentstyle[psfig,sprocl]{article}
\def\Journal#1#2#3#4{{#1} {\bf #2}, #3 (#4)}
\def\PLB{{\em Phys. Lett.}  B}

\def\PRD{{\em Phys. Rev.} D}

\def\be{\begin{equation}}
\def\ee{\end{equation}}
\def\bea{\begin{eqnarray}}
\def\eea{\end{eqnarray}}

\def\ell{l}

\def\bqll{$b \rightarrow q \ell^+ \ell^- $ }

\def\bxqll{$B \rightarrow X_q \ell^+ \ell^- $ }
\def\bxsll{$B \rightarrow X_s \ell^+ \ell^- $ }
\def\bxdll{$B \rightarrow X_d \ell^+ \ell^- $}

\def\absvcb{\left| V_{cb} \right|}

\def\s{\hat{s}}
\def\u{\hat{u}}

\def\c910{C_9^{\mbox{eff}} \mp C_{10}}

\def\mc{\hat{m}_c}

\def\mq{\hat{m}_q}
\def\lo{\hat{\lambda}_1}
\def\lt{\hat{\lambda}_2}

\def\Vuq{V_{uq}}
\def\Vub{V_{ub}}
\def\Vtq{V_{tq}}
\def\Vtb{V_{tb}}
\def\Vcq{V_{cq}}
\def\Vcb{V_{cb}}
\def\dBs{$ {dB \over ds} [\mbox{\bxsll}] $}
\def\dBd{$ {dB \over ds} [\mbox{\bxdll}] $}
\def\loeq{\begin{eqnarray}
& &\frac{{\rm d}{\cal B}}{{\rm d}\s}= 
   2         \frac {{\vert V_{tq}^*
                    V_{tb}\vert}^2}{\absvcb^2}
      \; {\cal B}_0
		\left\{ 
                  \left[
   		\frac{2}{3} \u(\s,\mq) 
((1 - \mq^2)^2 + \s (1+\mq^2) -2 \s^2)
\right. \right.
\nonumber \\  
   	& &
+	\frac{1}{3} (1 -4 \mq^2 + 6 \mq^4 -4 \mq^6 + \mq^8
-\s 
+ \mq^2 \s +
\mq^4 \s - \mq^6 \s 
\nonumber \\
& &
 -3 \s^2 -2 \mq^2 \s^2 -3 \mq^4 \s^2 
+ 5 \s^3 +5 \mq^2 \s^3-2 \s^4 ) \frac{\lo}{ \u(\s,\mq)}  
\nonumber \\
  & &       + \left( 1 -8 \mq^2 + 18 \mq^4 -16 \mq^6 + 5 \mq^8 -\s 
-3 \mq^2 \s + 9 \mq^4 \s -5 \mq^6 \s -15 \s^2
\right. \nonumber \\
& & 
\left. \left.-18 \mq^2 \s^2
-15 \mq^4 \s^2 + 25 \s^3
 + 25 \mq^2 \s^3 -10 \s^4 \right) \frac{\lt}{ \u(\s,\mq)}  
			\right] 
     		\left( |C_9^{\mbox{eff}}|^2 + |C_{10}|^2 \right)
		\nonumber \\
	&  &
            + 	\left[
	 	 \frac{8}{3} \u(\s, \mq) (2 (1+\mq^2)(1-\mq^2)^2-
                (1+14 \mq^2 +\mq^4) \s -(1+\mq^2) \s^2 )
		\right.
		\nonumber \\
  	& &  
          +	\frac{4}{3} (2 - 6 \mq^2 + 4 \mq^4 +4 \mq^6 -6 \mq^8+
 2 \mq^{10} 
\nonumber \\ & &
-5 \s -12 \mq^2 \s + 34 \mq^4 \s -12 \mq^6 \s -5 \mq^8 \s + 3 \s^2
\nonumber \\
& & 
 + 29 \mq^2 \s^2 + 29 \mq^4 \s^2 
+3 \mq^6 \s^2+ \s^3 -10 \mq^2 \s^3 +\mq^4 \s^3-\s^4-\mq^2 \s^4)
  \frac{\lo}{ \u(\s,\mq)} 
\nonumber \\
& &+ 4  \left(-6 + 2 \mq^2
 + 20 \mq^4 -12 \mq^6 - 14 \mq^8 +10 \mq^{10}
 + 
  3 \s+   16 \mq^2 \s +
 62 \mq^4 \s  \right. \nonumber \\ & & -56 \mq^6 \s
 -25 \mq^8 \s + 3 \s^2
 + 73 \mq^2 \s^2 + 101 \mq^4 \s^2 
+15 \mq^6 \s^2+ 5 \s^3
-26 \mq^2 \s^3+
\nonumber \\ & & \left.\left. 
5 \mq^4 \s^3 -5 \s^4-5 \mq^2 \s^4 \right) \frac{\lt}{ \u(\s,\mq)}
                \right] \frac{|C_7^{\mbox{eff}}|^2}{\s}
	         +      \left[
	8 \u(\s, \mq) \left((1-\mq^2)^2 \right.\right. \nonumber \\ & &
\left. -(1+\mq^2) \s \right)
                + 4 ( 1 - 2 \mq^2 +\mq^4  - \s-\mq^2 \s) 
\; \u(\s,\mq) \;  \lo  
\nonumber \\
   & & + 4 \left( -5 +30\mq^4-40 \mq^6 +15 \mq^8 -\s + 21
\mq^2 \s +25 \mq^4 \s -45 \mq^6 \s + 13 \s^2 
\right.  \nonumber \\
& & 
\left. \left. \left.
 + 22 \mq^2 \s^2
+45 \mq^4 \s^2 -7 \s^3-15 \mq^2 \s^3 \right)\frac{\lt}{ \u(\s,\mq)}    
			\right] Re(C_9^{\mbox{eff}}) \, C_7^{\mbox{eff}} 
		\right\}
\, .
\end{eqnarray}}
\begin{document}
\title{Determination of $|{V_{td} / V_{ts}}|$\\ 
within the Standard Model
}
\author{ C. S. Kim   }
\address{Department of Physics, Yonsei University, \\
Seoul 120-749, Korea}
\author{T. Morozumi \footnote{Talk is presented by T. Morozumi at BCONF97.}}
\address{Department of Physics, Hiroshima University, \\
1-3-1 Kagamiyama, Higashi Hiroshima, Japan 739}
\author{ A. I. Sanda }
\address{ Department of Physics, Nagoya University\\
    Chikusa-ku, Nagoya, Japan 464-01}
\maketitle\abstracts{
We propose a new method to extract the value of 
 $ |V_{td} / V_{ts}|$
from the ratio of the decay distributions 
$  {dBr \over ds} [\mbox{\bxdll}] $ and 
$  {dBr \over ds} [\mbox{\bxsll}] $. 
}
\section{A new method to extract $V_{td} / V_{ts}$ }
%#####################################################
The determination of the elements of CKM matrix is one of
the most important issues in the quark flavor physics.
In this  talk,   we study specifically  
$V_{td} / V_{ts}$. In the Standard Model with the
unitarity of CKM matrix, $V_{ts}$ is approximated by
$-V_{cb}$ which is directly measured by semileptonic B decays.
The unknown element $|V_{td}|$ can be extracted through
 $B_d - \bar{B_d}$ mixing.
However, in $ B_d -\bar{B_d}$ mixing the uncertainty of the hadronic
matrix elements prevents us to extract the element of CKM with
good accuracy.  
In this letter, we propose another method  to determine 
${|V_{td} /  V_{ts}|}$ with the decay distributions
$\mbox{\dBs}$ and  $\mbox{\dBd}$, where $s$
is invariant mass-squared of $l^+ l^-$. 
The idea stems from the fact that 
in the decays $\mbox{\bxqll} (q=d,s)$, the short distance (SD)
contribution  comes from the top quark loop diagrams 
and the long distance (LD) contribution  comes 
from the decay chains due to charmonium states.
The former amplitude is proportional to 
${V_{tq}}^* V_{tb}$ and the latter proportional to
${V_{cq}}^* V_{cb}$. If the invariant mass squared of $l^+
l^-$ is away from the peaks of the  charmonium resonances ($J /\psi$ and
$\psi'$), the SD contribution is dominant, while on the peaks
of the resonances 
the LD contribution is dominant$^{1}$. % \cite{lms}. 

In  SU(3) limit $m_d=m_s$, 
the ratio of the distributions 
\begin{equation} 
{dR \over ds} \equiv \mbox{\dBd} / \mbox{\dBs} 
\end{equation}
 becomes $|{V_{td} \over V_{ts}}|^2 $, if
$s$ is away from  the peak of  resonance $J/\psi$.
On the peak, the ratio becomes $|{V_{cd} \over V_{cs}}|^2=
\lambda^2 $, where $\lambda$
is Cabibbo angle, $sin \theta_c$. In the intermediate region, there is
characteristic interference between the LD contribution and 
the SD contribution, which requires the detailed study of
the distributions. To summarize our prediction in the SU(3) limit: 
\begin{eqnarray}
{1 \over \lambda^2}{dR(s) \over ds} & \rightarrow&  (1-\rho)^2+\eta^2
   \quad (s \rightarrow 1 \rm{GeV}^2), \nonumber \\
&\rightarrow & \qquad 1  \qquad  \qquad  ( s
\rightarrow  {m_{_{J/\psi}}}^2) . 
\end{eqnarray}
In the following, we present the explicit formulae for the
distribution as well as the numerical results for the ratio.

%#######################################################
\section{ The differential decay rates and the ratio }
%########################
Following the notation of Ref. [2], the amplitude of 
$ \mbox{\bqll} (q=d,s) $ can be written as$^3$  %, \cite{ks} 
\begin{eqnarray} 
	{\cal M (\mbox{\bqll})} & = & 
	\frac{G_{_F} \alpha}{\sqrt{2} \pi} \, V_{tq}^\ast V_{tb} \, 
	\left[ \left( C_{9q}^{\mbox{eff}} - C_{10} \right) 
		\left( \bar{q} \, \gamma_\mu \, L \, b \right)
		\left( \bar{l} \, \gamma^\mu \, L \, l \right) 
                \right. \nonumber \\
        & & \left.
                \; \; \; \; \; \; \; \; \;
                \; \; \; \; \; \; \; \; \;
		+ \left( C_{9q}^{\mbox{eff}} + C_{10} \right) 
		\left( \bar{q} \, \gamma_\mu \, L \, b \right)
		\left( \bar{l} \, \gamma^\mu \, R \, l \right)  
		\right. \nonumber \\
	& & \left. 
                         \; \; \; \; \; \; \; \; \;
	- 2 C_7^{\mbox{eff}} \left( \bar{q} \, i \, \sigma_{\mu \nu} \, 
		\frac{q^\nu}{q^2} (m_q L + m_b R) \, b \right)  
		\left( \bar{l} \, \gamma^\mu \, l \right) 
		\right] 
	\label{eqn:hamiltonian}
\end{eqnarray} where $C_{9q}^{\mbox{eff}}$ is given by,
\begin{equation}
C_{9q}^{\mbox{eff}} (\hat{s}) \equiv C_9
%\eta({\hat{s}}) +
\left\{ 1 + \frac{\alpha_s(\mu)}{\pi}
                \omega(\s) \right\} +
{Y_{SD}}^q (\hat{s})
 + {Y_{LD}}^q (\hat{s}),
\end{equation}
and  $q^\nu$ is the sum of the four momentum of $ l^+ l^-$, 
$s=q^2$ and  $\hat{s}=s / {m_b}^2$.
The function ${Y_{SD}}^q (\hat{s})$ is the one-loop matrix element of $O_9$,
and  ${Y_{LD}}^q (\hat{s})$ is the LD contributions 
due to the vector mesons  $J/\psi$, $\psi^\prime$ 
and higher resonances. The function $\omega(\hat{s})$ represents the
$O(\alpha_s)$ correction from the one-gluon exchange in the matrix element of
$ O_9$. The function ${Y_{SD}}^q$ can be found in the literature$^3$, which 
is  written as
\begin{eqnarray}
         {Y_{LD}}^q(\s) & = & 
                      \frac{3}{\alpha^2} \kappa \,
                \left( -{ \Vcq^* \Vcb \over \Vtq^* \Vtb}  C^{(0)}
                  -{\Vuq^* \Vub \over \Vtq^* \Vtb} ( 3 \, C_3
                + C_4 + 3 \, C_5 + C_6 )\right) \nonumber \\
		& & \times \sum_{V_i = \psi(1s),..., \psi(6s)}
	\frac{\pi \, \Gamma(V_i \rightarrow l^+ l^-)\, M_{V_i}}{
		{M_{V_i}}^2 - \s \, {m_b}^2 - i M_{V_i} \Gamma_{V_i}} .
                \label{LDeq} %\nonumber    %\\
               \label{eqn:yld}
%        \eta(\s) & = & 1 + \frac{\alpha_s(\mu)}{\pi}
%                \omega(\s) ~,\label{eqn:yld}
\end{eqnarray}
%%%%%%%%%%%%%%%%%%%%%%%%%%%%%%%%%%%%%%%%%%%%%%%%%%%
In Eq. (\ref{eqn:yld}),
$C^{(0)} \simeq  3 C_1^{(0)} + C_2^{(0)} \simeq  0.38 (0.12)$
while  $ 3 C_3^{(0)} + 
		C_4^{(0)} + 3 \, C_5^{(0)} + C_6^{(0)}
\simeq -0.0014(-0.0035)$ for $\mu=5GeV(2.5GeV)$.
Therefore the second term in $Y_{LD}$ can be neglected.
The parameter $\kappa$ is introduced so that the correct
value for the branching fraction of  the decay chain
$ B \rightarrow X_q J/{\psi} \rightarrow X_q l^+ l^-$ is
reproduced. We choose $\kappa C^{(0)} =0.88$.   
The differential decay rate for the inclusive semileptonic
decay has been calculated for $\mbox{\bxsll}$ including the 
$ {\Lambda_{QCD}}^2 / {m_b}^2 $ power corrections.
The result can be translated into $\mbox{\bxqll}$. 
\loeq
In the above expression,
the branching ratio is normalized by 
${\cal B}_{sl}$ for  decays $B \to (X_c,X_u) \ell
\nu_\ell$. We separately write  a combination of the CKM factor 
$\frac {{\vert V_{tq}^* V_{tb}\vert}^2}{\absvcb^2}$      
due to top quark loop from the normalization factor  ${\cal B}_0$. 
The normalization constant ${\cal B}_0$ is
\begin{equation}
        {\cal B}_0  \equiv
               {\cal B}_{sl} \frac{3 \, \alpha^2}{16 \pi^2} 
                   \frac{1}{f(\mc) \kappa(\mc)}
                \, ,
\label{eqn:seminorm}
\end{equation}
where $f \simeq 0.54 $ is a phase space factor,  and  $\kappa(\mc)
\simeq 0.85 $ includes $O(\alpha_s)$ 
QCD correction and $1 /m_b^2$ corrections. 
The differential decay rate is not a simple parton model
result. It contains non-perturbative $1 / {m_b}^2$ power 
corrections and are denoted by the terms proportional to
$ \hat{\lambda}_1 \simeq -0.0087$ and $ \hat{\lambda}_2
\simeq 0.0052 $. Because of the smallness of these
parameters, 
the correction is small compared with the
leading parton model result for  the region of $ s <<
(m_b-m_q)^2$, $ (q=d,s)$.
%########.################

\section{\bf Numerical Results and Summary}
%The limit  on $|V_{td}|$ and $|V_{ub}|$ 
%and the extraction of $(1-\rho)^2 + \eta^2 $ }
%#############################################
 From the present experimental value of $ \Delta M_d=
0.474 \pm 0.031 (ps^{-1}) $, and from the analysis on 
$ B \rightarrow X_u l
\bar{\nu}$, we obtain the limits on
$ (1-\rho)^2 + \eta^2$  
and   $ \sqrt{\rho^2 + \eta^2}$:
\begin{eqnarray}
(1-\rho)^2 + \eta^2&=&
0.85(1 \pm 0.15) \left\{ 0.2 / ({B_B}^{1/2} F_B({\rm GeV})) 
\right \}^2 \nonumber \\
&=&[0.59 \pm 0.09 , 1.3 \pm 0.20 ] \nonumber \\ 
\sqrt{\rho^2 + \eta^2} &=& 0.36 \pm 0.09,
\end{eqnarray}
where we use the following  values,
$\eta_{QCD}=0.55$, $m_t=175(GeV)$
$ A \lambda^2 =0.041(\pm 0.003)$ and  $ \lambda=0.2205, F_B=0.2 \pm 0.04$.
Within  the limits, we show how  $ {1 \over \lambda^2}{dR \over ds} $ 
depends on $ (1-\rho)^2 + \eta^2$ and $\phi_1=
arg  (- { {V_{cd}V_{cb}^*} \over  {V_{td}V_{tb}^*}}) $.

%where ${B_B}^{1/2} F_B = 0.2 \pm 0.04 (\rm{GeV}) $.  
%In Fig.1, we show the limits in  the plane of
%$(\phi_1, (1-\rho)^2+\eta^2)$.
In Fig.1, we show the ratio  $ {1 \over \lambda^2}{dR(s) \over ds} $ 
for two  sets of values for $(\phi_1, (1-\rho)^2+\eta^2)$,
one of which $(1-\rho)^2+\eta^2$ is minimum and the other
is maximum. The ratios for the CP conjugate process 
$ {\bar B} \rightarrow {\bar X_q} l^+ l^- $ are  also shown. 
To summarize the numerical results:
\begin{itemize}
\item{ 
The ratio at low invariant mass region $ \simeq 1
(GeV^2)$ is very near to the value of $(1-\rho)^2 + \eta^2$,
while on the peak of the resonance $J/\psi$, the ratio
is almost 1. 
%\item{ The effect of SU(3) breaking is small.}
The value at low invariant 
mass region does not  depend  so much on whether
the decaying particles are 
$B$ or $ \bar B$.}
\item{
For more details on this work, please see Ref. [4].}
\end{itemize}

%###############################################
%\subsection{Limitations on the Placement of Tables,
%Equations and Figures}\label{sec:plac}
%\begin{figure}
%\rule{5cm}{0.2mm}\hfill\rule{5cm}{0.2mm}
%\vskip 2.5cm
%\rule{5cm}{0.2mm}\hfill\rule{5cm}{0.2mm}
%\psfig{figure=filename.ps,height=1.5in}
%\caption{Radiative (off-shell, off-page and out-to-lunch) SUSY Higglets.
%\label{fig:radish}}
%\end{figure}
%%%%%%%%%%%%%%%%%%%%%%%%%%%%%%%%%%%%%
%      Figures
%
%\begin{figure}[htb]
%\begin{center}
%\leavevmode\psfig{file=rb.ps,width=12cm}
%\end{center}
%\caption[]{ The limit on $(1-\rho)^2+\eta^2$ and   $ \phi_1(\beta)$
%The horizontal axis corresponds to $\phi_1(\beta)$ and the unit is 
%degree. The vertical axis corresponds to
%$(1-\rho)^2+\eta^2$.
%The dashed lines are obtained from the central values
%of $|V_{ub}|$ and $|V_{td}|$. }
%\label{fig:hillerfig1}
%\end{figure}
%%%%%%%%%%%%%%%%%%%%%%%%%%%%%%%%%%%%%%%
\begin{figure}[htb]
\begin{center}
\leavevmode\psfig{file=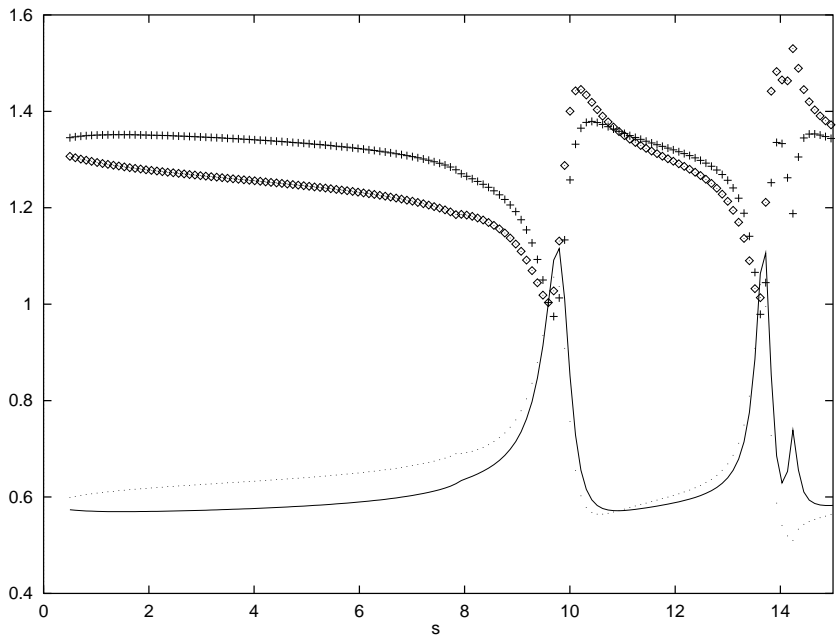,width=12cm,angle=-90}
\end{center}
\caption[]{
The ratio $ {1 \over \lambda^2}{dR(s) \over ds} $ 
versus $s$ (GeV$^2$)
with two different values of $(1-\rho)^2+\eta^2$.
The solid curve corresponds to 
$(\phi_1, (1-\rho)^2+\eta^2) =  (20,0.59)$, and the boxed curve   
corresponds to  $(20 ,1.33)$.  
The ratio for CP conjugate process 
$ {\bar B} \rightarrow {\bar X_q} l^+ l^- $
is denoted by the crossed curve  for $(1-\rho)^2+\eta^2=1.33$,
and by the dotted curve for $(1-\rho)^2+\eta^2=0.59.$
}
\label{fig:hillerfig1}
\end{figure}
%%%%%%%%%%%%%%%%%%%%%%%%%%%%%%%%%%%%%%%%%%%%%%%%%%%%%%%%
\section*{Acknowledgments}
We would like to thank the organizer of BCONF97.
%T. M. would like to thank G. Hiller for correspondence
%on $1/ mb^2$ corrections.
The work of T.M. is supported by 
Monbusho Grant, No. $08044089$.
 The work of C.S.K. is, in part, supported by
BSRI-97-2425, and the work of A.I.S. was, in part, supported by
Daiko Foundation. 

\end{document}